# Breaking the Loss Limitation of On-chip High-confinement Resonators

Xingchen Ji[1,2], Felippe A.S. Barbosa[2,*], Samantha P. Roberts[2], Avik Dutt[1,2], Jaime Cardenas[2,†], Yoshitomo Okawachi[4], Alex Bryant[5], Alexander L. Gaeta[4] and Michal Lipson[2]

[1]*School of Electrical and Computer Engineering, Cornell University, Ithaca, New York 14853, USA*

[2]*Department of Electrical Engineering, Columbia University, New York, New York 10027, USA*

[3]*School of Applied and Engineering Physics, Cornell University, Ithaca, New York 14853, USA*

[4]*Department of Applied Physics and Applied Mathematics, Columbia University, New York, New York 10027, USA*

[5]*School of Materials Science and Engineering, Georgia Institute of Technology, Atlanta, GA 30332, USA*

[*]*currently at Instituto de Física, Universidade Estadual de Campinas, 13083-970 Campinas, SP, Brasil*

[†]*currently at The Institute of Optics, University of Rochester, Rochester, NY 14627, USA*

**On-chip optical resonators have the promise of revolutionizing numerous fields including metrology and sensing; however, their optical losses have always lagged behind their larger discrete resonator counterparts based on crystalline materials and flowable glass. Silicon nitride ($Si_3N_4$) ring resonators open up capabilities for optical routing, frequency comb generation, optical clocks and high precision sensing on an integrated platform. However, simultaneously achieving high quality factor and high confinement in $Si_3N_4$ (critical for nonlinear processes for example) remains a challenge. Here, we show that addressing surface roughness enables us to overcome the loss limitations and achieve high-confinement, on-chip ring resonators with a quality factor ($Q$) of 37 million for a ring with 2.5 μm width and 67 million for a ring with 10 μm width. We show a clear systematic path for achieving these high quality factors. Furthermore, we extract the loss limited by the material absorption in our films to be 0.13 dB/m, which corresponds to an absorption limited $Q$ of at least 170 million by comparing two resonators with different degrees of confinement. Our work provides a chip-scale platform for applications such as ultra-low power frequency comb generation, high precision sensing, laser stabilization and sideband resolved optomechanics.**

Low propagation loss silicon nitride ($Si_3N_4$) ring resonators are critical for a variety of photonic applications such as efficient and compact on-chip optical routing [1,2], low threshold frequency combs [3-5], optical clocks [6] and high precision sensing [7-11]. High confinement is critical for tailoring the waveguide dispersion to achieve phase matching in nonlinear processes as well as for tighter bends in large scale photonic systems. A microresonator's quality factor ($Q$) is extremely sensitive to losses. To date, ultra-high $Q$'s have been demonstrated only in low confinement large mm-scale resonators based on platforms such as polished calcium fluoride ($CaF_2$), magnesium fluoride ($MgF_2$) or flowable silica glass [12-16] with typical cross sectional mode field diameter much larger than the wavelength. Spencer *et al.* have recently demonstrated ring resonators with a high $Q$ of up to 80 million using extremely thin (40 nm) $Si_3N_4$ films [17], which can be useful for narrowband filtering or building reference cavities for laser stabilization. However, they suffer from highly delocalized optical modes and millimeter-scale bending radii, making it challenging to use these thin film ring resonators for compact photonic routing or nonlinear applications requiring dispersion engineering. The highest repeatable intrinsic $Q$ in high confinement $Si_3N_4$ ring resonators reported to date is 7 million [18].

In this work, we show that surface roughness, rather than absorption from the bulk material, plays a major role in the loss limitations of $Si_3N_4$ therefore enabling a path for achieving ultra low loss devices by simply addressing surface quality. Absorption loss is mainly due to O-H bonds in $SiO_2$, and N-H and Si-H bonds in $Si_3N_4$ [19]. Scattering loss comes primarily from the interaction of light with the roughness of all the surfaces in a high confinement waveguide. Mode simulations show that light propagating in the waveguide significantly interacts and scatters from both the patterned sidewalls and the top and bottom surfaces (See Fig. 1c). Several groups have been working on reducing losses by improving the bulk material properties to achieve high $Q$ [20-22]. However, to date it has not been clear whether surface interactions or material absorption is the main source of the high loss in the integrated platform.

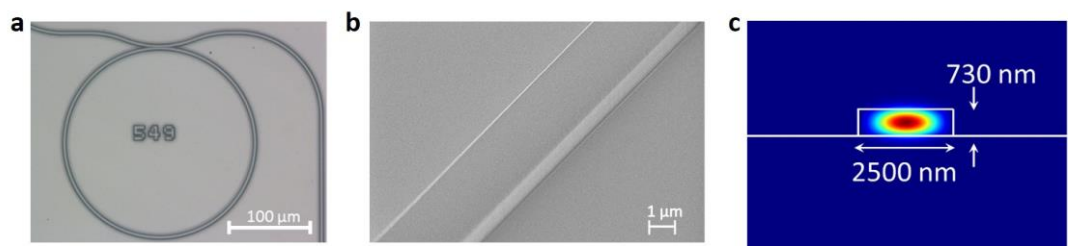

**Figure 1 | Microscope images and mode simulation of fabricated devices. a**. Top view optical microscope image of a 115 µm radius ring resonator. **b**. Scanning electron microscopy image of a fabricated waveguide with smooth surfaces. **c**. Mode simulation of 730 nm tall and 2500 nm wide waveguide showing that the mode is highly confined in the geometry we have chosen.

In order to reduce surface scattering from the sidewalls, we minimize the roughness introduced during the pattern transfer step of the processing by eliminating in-situ polymer formation typical in dry etching processes. Standard waveguide fabrication methods consist of patterning a masking layer, typically photoresist or electron-beam resist, and transferring this pattern into the photonic waveguide device layer using some form of plasma etching [23,34]. Polymer formation is a common by-product of plasma etching. In-situ polymer deposition passivates the sidewalls and enables anisotropic etching with vertical sidewalls desirable for rectangular waveguide fabrication [25,26]. It also enables pattern transfer to thick waveguide device layers by enhancing selectivity between the mask layer and the films. The polymer formed during this process while critical for surface passivation and anisotropic etching often leaves residue on the sidewalls which introduces sidewall roughness. This roughness adds to the one introduced by the lithography itself [27,28]. Since the roughness is generally on the order of nanometers, it usually introduces negligible loss, however, it becomes significant in the high $Q$ regimes that we are aiming for here [19,29]. Trifluoromethane ($CHF_3$) and oxygen ($O_2$) gases are widely used as standard etchants in $Si_3N_4$ fabrication and this etching chemistry is always accompanied by polymer residue left on the sidewalls [30,31]. In order to reduce this polymer residue on sidewalls, we used a higher oxygen flow to remove in-situ polymer formation, since oxygen reacts with polymer residue to form carbon monoxide (CO) and carbon dioxide ($CO_2$). Oxygen also reacts with the photoresist which is generally used in standard etching as the mask to transfer patterns. As a result, higher oxygen flow decreases the etching selectivity, degrading the ability to transfer patterns. To compensate for this effect, we use a silicon dioxide hard mask instead of photoresist to maintain the ability to transfer waveguide patterns while eliminating in-situ polymer formation on the sidewalls using higher oxygen flow. Nitrogen is also added to increase the nitride selectivity over oxide [32,33].

In contrast to standard silicon-based waveguides with losses on the order of 1 dB/cm [34-36] where the sidewall roughness plays the major role in inducing scattering loss, in ultra low loss $Si_3N_4$ the top surface roughness also plays a major role. Typically roughness on the top and/or bottom surfaces has not attracted much attention due to the facts that the sidewall roughness was quite significant and many of the previous studies have relied on polished wafers or oxidized wafers from silicon photonics. Here we focus on reducing scattering loss from the top surface since the $Si_3N_4$ films are deposited using low-pressure chemical vapor deposition (LPCVD), which are not as inherently smooth as polished single-crystal wafers or oxidized wafers. The bottom surface roughness is not addressed here since its roughness, governed by thermal oxidation, is lower than the one governed by the $Si_3N_4$ deposition (see AFM scans in supplementary section).

In order to reduce scattering from the top surfaces, we reduce the roughness by chemical mechanical polishing (CMP) the $Si_3N_4$ after the deposition (as shown in Fig. 2d). The atomic force microcopy (AFM) scans before and after the polishing step are shown in Fig. 2. The root mean squared (RMS) roughness is decreased from 0.38 nm to 0.08 nm (AFM scans of different CMP $Si_3N_4$ films are shown in the supplementary section).

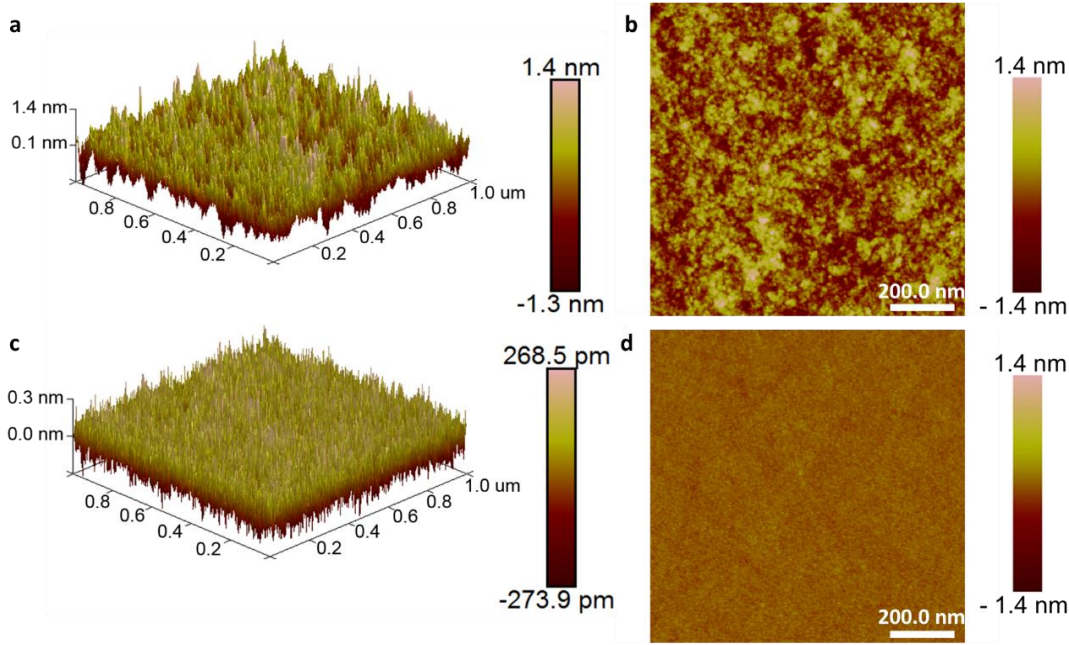

**Figure 2 | AFM measurement of Si$_3$N$_4$ top surface. a**. 3D AFM scan of Si$_3$N$_4$ top surface before CMP with RMS roughness of 0.38 nm. **b**. 2D image of Si$_3$N$_4$ top surface before CMP and scaled to -1.4 – 1.4 nm with RMS roughness of 0.38 nm. **c**. 3D image of Si$_3$N$_4$ top surface after CMP with RMS roughness of 0.08 nm. **d**. 2D image of Si$_3$N$_4$ top surface after CMP and scaled to -1.4 – 1.4 nm with RMS roughness of 0.08 nm. Note the different scale bars on (a) and (c).

To further decrease the loss, we apply multipass lithography to reduce line edge roughness known to contribute to scattering loss [37-39]. Electron beam (e-beam) lithography, extensively used for pattering optical waveguides, creates a line edge roughness which introduces extra roughness to the sidewalls. During e-beam lithography, any instability, such as beam current fluctuations, beam jitter, beam drift, stage position errors and mechanical vibrations, can generate statistical errors which result in extra line edge roughness in the patterns which will add roughness to the sidewalls. The principle of multipass lithography [38,39] consists of exposing the same pattern multiple times at a lower current to reduce line edge roughness by averaging statistical errors.

We measure an intrinsic $Q$ of 37 ± 6 million in high confinement Si$_3$N$_4$ ring resonators using the techniques described above. Mode splitting, commonly observed in ultra-high $Q$ system such as whispering-gallery-mode microresonators [40-42], is induced due to light backscattering from fabrication imperfections or surface roughness. When the $Q$ is high and the mode is highly confined, extremely small defects or roughness can induce a visible splitting. We measured the transmission of four sets of fabricated ring resonators: 1) using the standard process reported in Ref. 18 (Fig. 3a), 2) using our optimized etch process but without CMP and without multipass lithography (Fig. 3b), 3) using both the optimized etch recipe and CMP but without multipass lithography (Fig. 3c), and 4) using all the techniques including the optimized etch recipe, surface smoothing technique and multipass lithography. All the rings have a radius of 115 μm, a height of 730 nm and a width of 2500 nm, and are coupled to a waveguide of the same dimensions. The transmission spectra and the linewidth of the resonator (FWHM) are measured using a laser scanning technique. We launch light from a tunable laser source which is then transmitted through a fiber polarization controller and coupled into our device via an inverse nanotaper [43] using a lensed fiber. We collect the output of the ring resonator through another inverse nanotaper and an objective lens. We monitor the output on a high speed InGaAs photodetector. The frequency of the laser is measured using a wavemeter with a precision of 0.1 pm and the laser detuning is calibrated by monitoring the fringes of a reference fiber based Mach-Zehnder interferometer with a known free spectral range (FSR).

Figure 3 shows the measured transmission spectra of different ring resonators. The measured intrinsic $Q$'s, estimated by measuring the transmission [44,45] for rings using the different fabrication processes a-d described above, are 5.6 ± 0.7 million, 16.2 ± 2.9 million, 28 ± 4.7 million and 37 ± 6 million which correspond to propagation losses of 5.2 ± 0.6 dB/m, 1.8 ± 0.3 dB/m, 1.1 ± 0.2 dB/m and 0.8 ± 0.1 dB/m respectively [46]. Note that these estimated propagation losses are upper bounds on the losses in straight waveguides since in a ring the optical mode interacts more strongly with the sidewalls due to bending.

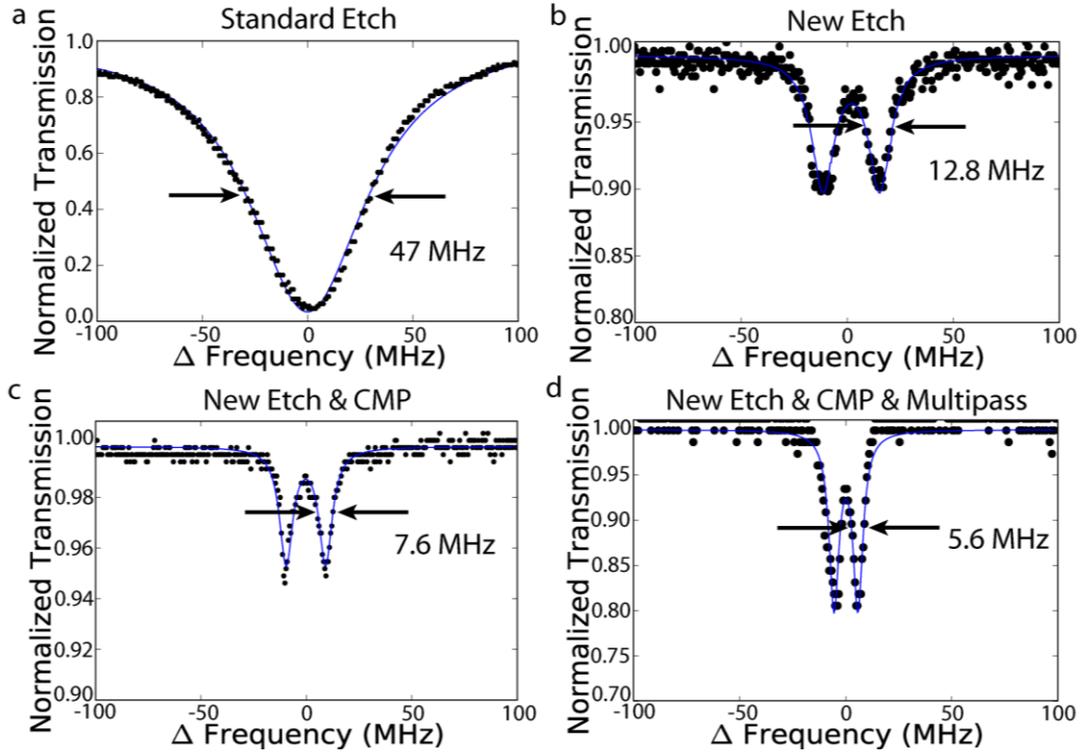

**Figure 3 | Normalized transmission spectra of ring resonators fabricated using different processes. a**. Device fabricated using the standard process reported in Ref. 18 with a measured full width half maximum (FWHM) of 47 MHz. **b**. Device fabricated using the optimized etch process but without our new surface smoothing technique and multipass lithography with a measured full width half maximum (FWHM) of 12.8 MHz. **c**. Device fabricated using both the optimized etch recipe and surface smoothing techniques but without multipass lithography with a measured full width half maximum (FWHM) of 7.6 MHz. **d**. Device fabricated using all the techniques including the optimized etch recipe, surface smoothing technique and multipass lithography with a measured full width half maximum (FWHM) of 5.6 MHz.

In order to illustrate the importance of simultaneous high $Q$ and high confinement ring resonators we demonstrate a strong decrease in the threshold for optical parametric oscillation down to sub-milliwatts with the decrease of optical losses. To determine the threshold for parametric oscillation, we measured the output power in the first generated four-wave-mixing (FWM) sideband for different pump powers. Fig. 4a shows the data for a device pumped at the resonance near 1557 nm with a loaded $Q$ of 35 million. The average threshold power is 330 ± 70 µW, comparable to the theoretically estimated threshold power of 206 µW using the expression [47,48]

$$P_{th} \approx 1.54(\frac{\pi}{2})\frac{Q_C}{2Q_L} \cdot \frac{n^2 V}{n_2 \lambda Q_L^2} \quad (1)$$

where $\lambda$ is the pump wavelength, $n$ is the linear refractive index, $n_2$ is the nonlinear refractive index which equals to $2.4 \times 10^{-19}$ m$^2$/W [49], $V$ is the resonator mode volume, $Q_c$ and $Q_L$ are the coupling and loaded quality factors of the resonators. This is the lowest and the first sub-milliwatt power threshold parametric oscillation in planar nonlinear platforms [50-54] reported to the best of our knowledge (Comparisons shown in Table 1). In addition, this threshold power is close to the lowest threshold reported in ultra-high $Q$ microresonators such as CaF$_2$ [55] and flowable silica glass [4]. We also measure and plot the thresholds for rings with various loaded quality factors in Fig. 4b. The threshold powers follow the theoretically predicted trend of being inversely proportional to $Q_L^2$.

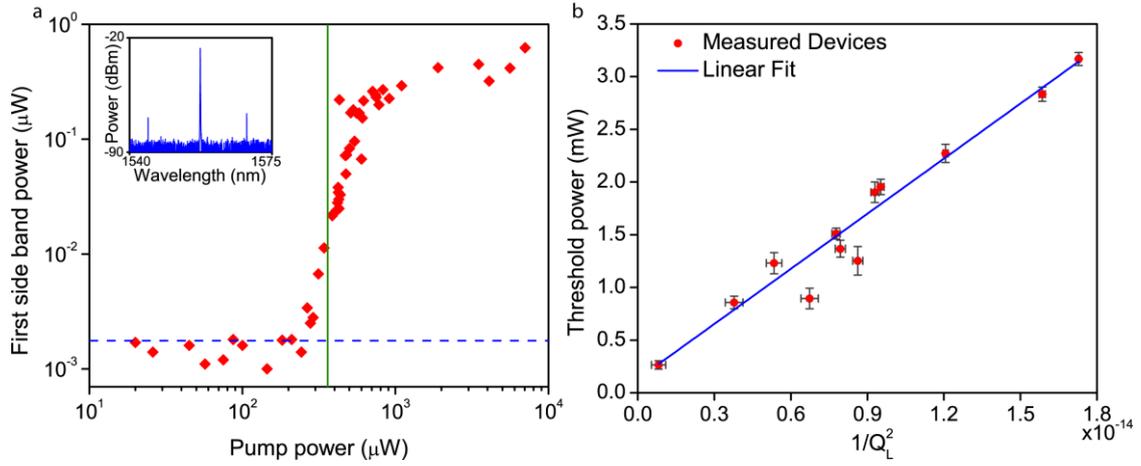

**Figure 4 | Oscillation threshold decrease with decrease of losses. a**. The output power in the first generated mode as a function of the pump power. In this device, parametric oscillation occurs for a pump power of 330 ± 70 µW (indicated by the solid green vertical line). Note that the first band appears more than one free spectral range (FSR) away from the pumped resonance. **b**. Measured threshold power for micro-resonators with different fabrication processes as a function of the loaded quality factor ($Q_L$). Threshold powers approximately follow the theoretically predicted trend of being inversely proportional to $Q_L^2$.

**Table 1. Comparison of Planar Nonlinear Platforms for On-Chip Frequency Comb Generation at Telecom Wavelengths**

| Platform | $n$ | $n_2$ (m$^2$/W) | Cross sections(µm$^2$) | FSR (GHz) | Q | $P_{th}$(mW) |
|---|---|---|---|---|---|---|
| Hydex [50] | 1.7 | $1.2 \times 10^{-19}$ | $1.45 \times 1.50$ | 200 | $1 \times 10^6$ | 50 |
| AlN [51] | 2.1 | $2.3 \times 10^{-19}$ | $0.65 \times 3.50$ | 435 | $8 \times 10^5$ | 200 |
| Diamond [52] | 2.4 | $8.2 \times 10^{-20}$ | $0.95 \times 0.85$ | 925 | $1 \times 10^6$ | 20 |
| Al$_{0.17}$Ga$_{0.83}$As [53] | 3.3 | $2.6 \times 10^{-17}$ | $0.32 \times 0.62$ | 995 | $1 \times 10^5$ | 3 |
| Si$_3$N$_4$ [54] | 2.0 | $2.5 \times 10^{-19}$ | $0.60 \times 3.00$ | 25 | $1.7 \times 10^7$ | 5.6 |
| Si$_3$N$_4$ (our present work) | 2.0 | $2.5 \times 10^{-19}$ | $0.73 \times 2.50$ | 200 | $3.6 \times 10^7$ | 0.33 ± 0.07 |

Table 1 Parametric oscillation threshold power for different planar nonlinear platforms.

In order to extract the fundamental limit of achievable loss in silicon nitride waveguides, we compare the losses of two different structures which have different mode interactions with the sidewalls. We estimate the bulk absorption limitation in our $Si_3N_4$ films to be 0.13 ± 0.05 dB/m, which corresponds to an absorption-loss-limited $Q$ of at least 170 million. We fabricated two devices with waveguide widths of 2.5 microns and 10 microns on the same wafer to ensure that the fabrication processes are identical. Both rings have the same height of 730 nm and both of them are coupled to a waveguide of the same dimensions (730 nm x 2500 nm). Figure 5b and 5c show the measured transmission spectra for the rings with 10 um width in TE and TM polarization. The measured intrinsic $Q$ is 67 ± 7 million for the TE mode and 59 ± 12 million for the TM mode. At these ultra-high $Q$'s, one is operating near the limits of $Q$ that can be reliably estimated by scanning a laser across a resonance. Hence, we corroborate these $Q$ measurements by performing a cavity ring-down experiment for the TM mode. As shown in Fig 5d, the measured lifetime is 25.6 ± 1.3 ns which corresponds to an intrinsic $Q$ of 63 ± 3 million, consistent with our measurement of the $Q$ using a laser scanning technique. We estimate the fundamental loss limit given by the bulk absorption of $Si_3N_4$ in our films ($\alpha_{total\_absorption}$) by comparing the losses for the two structures extracted from the transmission measurements ($\alpha_{ring}$ ~ 0.79 ± 0.14 dB/m and $\alpha_{wide\_ring}$ ~ 0.43 ± 0.046 dB/m) and considering the absorption of the rings with narrower and wider waveguides to be:

$$\alpha_{ring} = \alpha_{total\_absorption} + \alpha_{top\_scatter} + \alpha_{bottom\_scatter} + \alpha_{sidewalls\_scatter} \qquad (2)$$

$$\alpha_{wide\_ring} = \eta_1 \alpha_{total\_absorption} + \eta_2 (\alpha_{top\_scatter} + \alpha_{bottom\_scatter}) + \eta_3 \alpha_{sidewalls\_scatter} \qquad (3)$$

$\eta_1, \eta_2, \eta_3$ are the factors that account for the interaction of the field with the waveguide core, the top and bottom surfaces, and sidewalls respectively for the wider waveguides relative to the narrower waveguide [56] and are calculated using FEM simulations (performed with COMSOL) to be 1.010, 1.002 and 0.138 respectively. $\alpha_{top\_scatter}$ ~ 0.0066 dB/m (± 0.001 dB/m) and $\alpha_{bottom\_scatter}$ ~ 0.2408 dB/m (± 0.02dB/m) are the loss due to scattering at the top and bottom interfaces estimated from Payne-Lacey model [57] that relates scattering loss to the surface's rms ($\sigma$) roughness and correlation length ($L_c$) which are both extracted from the AFM measurements. The scattering losses due to the sidewalls $\alpha_{sidewalls\_scatter}$ and the bulk loss are then extracted to be 0.41 ± 0.05 dB/m and 0.13 ± 0.05 dB/m. Note that here we are assuming that both sidewalls have the same loss as they are experiencing the same ebeam and plasma etching conditions, and the loss in the oxide cladding is negligible when compared to the loss in the $Si_3N_4$ due to the high degree of confinement.

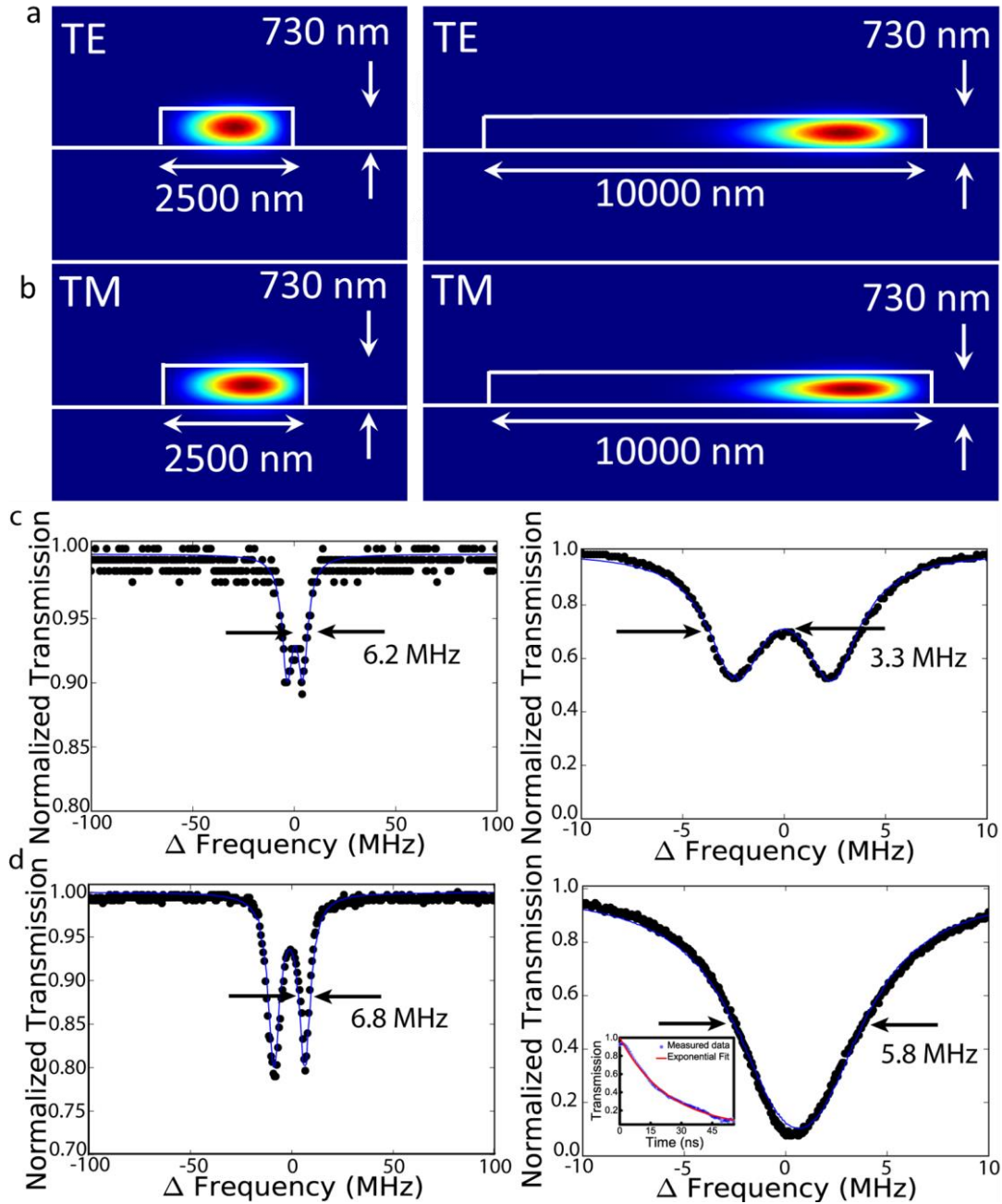

**Figure 5 | Mode simulation and normalized transmission spectra for ring resonators with different interaction strength with the sidewalls. a.** TE Mode profile of waveguides that are 2.5 µm and 10 µm wide and 730nm high. **b**. Same as a. but for TM. **c**. Measured normalized TE transmission spectra for the ring resonator composed of the 2.5 µm wide waveguide (left) with a measured full width half maximum (FWHM) of 6.2 MHz and the measured spectra for the ring resonator composed of the 10 µm wide waveguide (right) with a measured full width half maximum (FWHM) of 3.3 MHz in TE polarization using the optimized fabrication process. **d**. TM transmission spectra for the rings with narrower (left) and wider (right) waveguide with full width half maximum (FWHM) of 6.8 MHz and 5.8 MHz, respectively. . Inset shows the cavity ring-down measurement. The measured lifetime is extracted from the exponential fit to be 25.6 ± 1.3 ns.

In conclusion, we drastically and systematically reduced losses by using different methods for reducing the roughness from waveguide interfaces. These fabrication steps could not only enable one to achieve ultra-low loss in $Si_3N_4$ but also in other material platforms, independent of the geometry.

Moreover, we demonstrate optical parametric oscillation in an on-chip microresonator, with sub-milliwatt pump powers. We extract the absorption limited *Q* of the ring resonator to be at least 170 million, which indicates that we are still limited by the scattering loss, therefore providing a path for achieving ultra low loss resonators simply via addressing scattering loss. From our AFM measurements one possible path for further decreasing these scattering losses is by addressing the roughness at the bottom cladding/core interface generated by the thermal oxidation process. Our work provides an on-chip platform for devices with performances that could be comparable to the ones achieved in discrete large devices.

## Methods

**Device fabrication**

Starting from a virgin silicon wafer, a 4-um-thick oxide layer is grown for the bottom cladding. $Si_3N_4$ is deposited using low-pressure chemical vapor deposition (LPCVD) in steps. After $Si_3N_4$ deposition, CMP is applied to smooth the nitride films. After CMP, we deposit a $SiO_2$ hard mask using plasma enhanced chemical vapor deposition (PECVD). We pattern our devices with electron beam lithography while applying a multipass lithography technique. Ma-N 2403 resist was used to write the pattern and the nitride film was etched in an inductively coupled plasma reactive ion etcher (ICP RIE) using a combination of $CHF_3$, $N_2$, and $O_2$ gases. After stripping the resist and oxide mask, we anneal the devices at 1200°C in an argon atmosphere for 3 hours to remove residual N-H bonds in the $Si_3N_4$ film. We clad the devices with 500 nm of high temperature silicon dioxide (HTO) deposited at 800°C followed by 2.5 µm of $SiO_2$ using PECVD.

**Acknowledgments**

The authors gratefully acknowledge support from Defense Advanced Research Projects Agency under the Microsystems Technology Office Direct On-Chip Digital Optical Synthesizer program (N66001-16-1-4052) supervised by Dr. Robert Lutwak and the Air-Force Office of Scientific Research (FA9550-15-1-0303) supervised by Dr. Enrique Parra. This work was performed in part at the City University of New York Advanced Science Research Center and Cornell Nano-Scale Facility, a member of the National Nanotechnology Infrastructure Network, which is supported by the National Science Foundation (grant ECS-0335765). Xingchen is also grateful to the China Scholarship Council for financial support.

**Supplementary Information**

**1. AFM measurements of Si$_3$N$_4$ top surface using different CMP recipes**

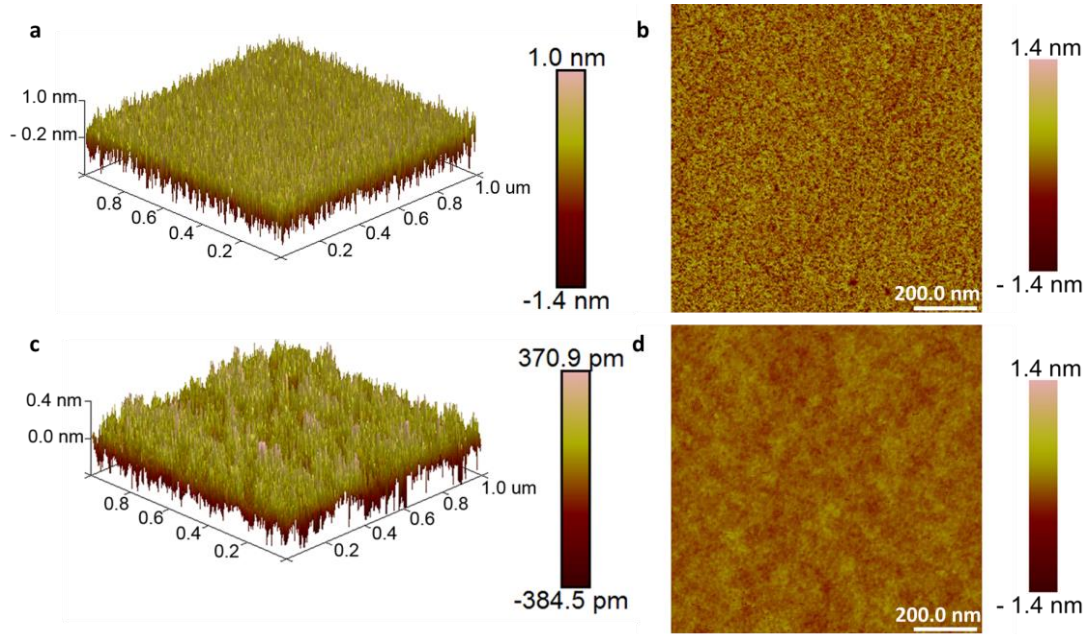

**FIG 1S| AFM measurements of Si$_3$N$_4$ top surface. a**. 3D AFM scan of Si$_3$N$_4$ top surface after CMP using a different polishing pad with RMS roughness of 0.32 nm. **b**. 2D image of Si$_3$N$_4$ top surface before CMP and scaled to -1.4 – 1.4 nm with a RMS roughness of 0.32 nm. **c**. 3D image of Si$_3$N$_4$ top surface after CMP using a different slurry with a RMS roughness of 0.11 nm. **d**. 2D image of Si$_3$N$_4$ top surface after CMP and scaled to -1.4 – 1.4 nm with a RMS roughness of 0.11 nm. Note the different scale bars on (a) and (c).

Our AFM measurements indicate that different polishing conditions can affect strongly the RMS roughness of the Si$_3$N$_4$ top surface. By comparing Fig. 1S a with Fig 2. from the main text, one can see that the pad selection has a significant effect on reducing Si$_3$N$_4$ top surface roughness. Comparing Fg.1S c with Fig 2. from the main text, one can see that the slurry selection also has a significant effect on roughness uniformity. While the RMS roughness of Si$_3$N$_4$ top surface is reduced, when compared with Fig 2. c, the roughness has more randomness. As a conclusion, both pad selection and slurry selection are important for reducing surface roughness.

**2. AFM measurements of oxidized Si wafer surface**

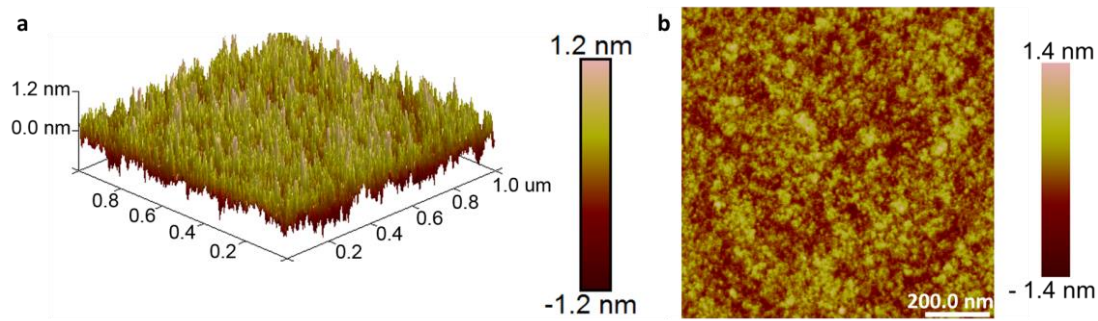

**FIG 2S | AFM measurement of oxidized Si wafer surface. a**. 3D AFM scan of oxidized Si wafer surface with a RMS roughness of 0.29 nm. **b**. 2D image of oxidized Si wafer surface before CMP and scaled to -1.4 – 1.4 nm with a RMS roughness of 0.29 nm. Note that in this work no CMP is performed on this bottom cladding/core interface surface.